\documentclass[11pt,superscriptaddress,aps,prd,preprint]{revtex4}
\usepackage[dvips]{graphicx}
\usepackage{amsfonts}
\usepackage{slashed}
\usepackage{amsmath}
\usepackage{amssymb}

\newcommand{\incps}[5]{\includegraphics[#2,#3][#4,#5]{#1}}

\begin{document}

\title{Lorentz-violating Euler-Heisenberg effective action}

\author{J. Furtado}
\affiliation{Instituto de F\'\i sica, Universidade Federal de Alagoas, 57072-270, Macei\'o, Alagoas, Brazil}
\email{jfurtado,tmariz@fis.ufal.br}

\author{T. Mariz}
\affiliation{Instituto de F\'\i sica, Universidade Federal de Alagoas, 57072-270, Macei\'o, Alagoas, Brazil}
\email{jfurtado,tmariz@fis.ufal.br}

\date{\today}

\begin{abstract}

In this work, we study the radiative generation of the Lorentz-violating Euler-Heisenberg action, in the weak field approximation. For this, we first consider a nonperturbative calculation in the coefficient $c_{\mu\nu}$, however, by assuming rotational invariance. Within this approach, we also obtain the results of the amplitude for the photon triple splitting. In the following, we take into account the perturbative approach, where $c_{\mu\nu}$ is treated as a insertion in the propagator and a new vertex. The partial results are in fact an expansion up to first order in $c_{\mu\nu}$ of the nonperturbative ones, with $c_{00}=\kappa$ and $c_{0i}=c_{ij}=0$. This suggest that the complete results obtained in the nonperturbative approach can be used in both treatments. 

\end{abstract}

\maketitle

\section{Introduction}\label{intro}

The Euler-Heisenberg effective action \cite{Heisenberg:1935qt,Schwinger:1951nm} has been widely investigated in the last decades, in various contexts (see \cite{Dunne:2012vv} and references therein). Nevertheless, studies related to the Lorentz-symmetry violation \cite{Colladay:1998fq,Colladay:1996iz,Kostelecky:2003fs} have been considered so far in the context of photon splitting \cite{Jacobson:2002hd,Adam:2002rg,Kostelecky:2002ue,Gelmini:2005gy,Kaufhold:2005vj,Brito:2011hc}. The effective theory used in these analyses was mainly the Lorentz-violating QED \cite{Kostelecky:2001jc}, a subset of the minimal standard-model extension \cite{Colladay:1998fq,Colladay:1996iz}. For an overview and references on this extended QED, see the review \cite{Bluhm:2005uj} and the incomplete list of works \cite{Jackiw:1999yp,Kostelecky:2001mb,Kostelecky:2002hh,Kostelecky:2009zp,Casana:2009dq,Gomes:2009wn,Casana:2010nd,Mariz:2010fm}.

In this paper, we are interested in studying the radiative generation of the Lorentz-violating Euler-Heisenberg action, from the CPT-even derivative term $i\bar\psi c_{\mu\nu}\gamma^\mu(\partial^\nu+ieA^\nu)\psi$ of the Lorentz-violating QED extension. The CPT-odd term $\bar\psi b_\mu \gamma^\mu\gamma_5\psi$ also generates nonlinear corrections to the Maxwell theory, however, these calculations well be presented in a next work. With the exact expressions for these results we will be able to calculate specific scattering amplitudes, e.g., the photon scattering in the electromagnet field of a nucleus (Delbr\"uck scattering) \cite{Jarlskog:1974tx,Milstein:1994zz} and the photon splitting in a strong magnetic field \cite{BialynickaBirula:1970vy,Adler:1970gg,Adler:1971wn,Adler:1996cja}, in order to numerically estimate the coefficients for the Lorentz violation.

Recently it has been argued that the photon-photon scattering (also a process calculated from the results of the Euler-Heisenberg action) can be observed at the Large Hadron Collider \cite{d'Enterria:2013yra} as an opportunity to discuss the noncommutative interactions \cite{Hewett:2000zp}, among others, and therefore also the Lorentz violation effects.

The structure of the paper is as follows. In Sec.~\ref{nonperturbative} we consider a nonperturbative calculation in the coefficient $c_{\mu\nu}$, by assuming rotational invariance in which $c_{\mu\nu}$ is reduced to the product of timelike vectors. We also recover the results of the amplitude for vacuum photon splitting, previously obtained in \cite{Kostelecky:2002ue}. In Sec.~\ref{perturbative} we take into account the perturbative approach, where the coefficient $c_{\mu\nu}$ is treated as a insertion in the propagator as well as a new vertex. Sec.~\ref{summary} contains a summary of out results.

\section{Nonperturbative approach}\label{nonperturbative}

The starting fermion Lagrangian that we are interested is given by
\begin{equation}\label{L}
{\cal L}_f=\bar\psi(i\tilde\partial_\mu\gamma^\mu- m - e\tilde A_\mu \gamma^\mu)\psi,
\end{equation}
where we have written $\tilde\partial_\mu=(g_{\mu\nu}+c_{\mu\nu})\partial^\nu$ and $\tilde A_\mu=(g_{\mu\nu}+c_{\mu\nu})A^\nu$. As we have mentioned above, in our nonperturbative approach the coefficient $c_{\mu\nu}$ is reduced to the product of two unit timelike vectors, i.e., $c_{\mu\nu}=\kappa u_\mu u_\nu$, where $u_\mu=(1,0,0,0)$ and $\kappa$ is the coefficient that determine the scale of Lorentz violation. Hence, the corresponding Feynman rules are the fermion propagator,
\begin{equation}
\raisebox{-0.4cm}{\incps{ferm.eps}{-1.5cm}{-.5cm}{1.0cm}{.5cm}}
 = \frac{i}{\slashed{\tilde p}-m}, 
\end{equation}
and the fermion-photon vertex,
\begin{equation}
\raisebox{-0.4cm}{\incps{vert.eps}{-1.5cm}{-.5cm}{1.0cm}{1.5cm}}
 = -ie\gamma^\mu,
\end{equation}
with $\slashed{\tilde p}=\tilde p_\mu\gamma^\mu$ and $\tilde p_\mu=((1+\kappa)p_0,p_i)$. Thus, the resulting effective action takes the form
\begin{eqnarray}\label{Seff}
S_{\rm eff}^{(4)} &=& \frac14 \int d^4x \int d^4k_1d^4k_2d^4k_3d^4k_4 \; e^{i(k_1+k_2+k_3+k_4)\cdot x} \nonumber\\
&&\times \frac16 G^{\mu_1\mu_2\mu_3\mu_4}(k_1,k_2,k_3,k_4) \tilde A_{\mu_1}(k_1)\tilde A_{\mu_2}(k_2)\tilde A_{\mu_3}(k_3)\tilde A_{\mu_4}(k_4),
\end{eqnarray}
where
\begin{eqnarray}\label{G}
G^{\mu_1\mu_2\mu_3\mu_4}(k_1,k_2,k_3,k_4) &=& 2\,T_1^{\mu_1\mu_2\mu_3\mu_4}(k_1,k_2,k_3,k_4) +  2\,T_2^{\mu_1\mu_2\mu_3\mu_4}(k_1,k_2,k_3,k_4) \nonumber\\
&&+ 2\,T_3^{\mu_1\mu_2\mu_3\mu_4}(k_1,k_2,k_3,k_4),
\end{eqnarray}
with
\begin{eqnarray}\label{T1}
\raisebox{-0.2cm}{\incps{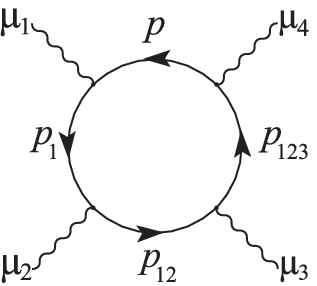}{0.0cm}{1.2cm}{3.5cm}{3.0cm}} &=& i\,T_1^{\mu_1\mu_2\mu_3\mu_4}(k_1,k_2,k_3,k_4) \nonumber\\ \nonumber\\
 &=& -(-ie)^4\int \frac{d^4p}{(2\pi)^4}{\rm tr}\frac{i}{\slashed{\tilde p}-m}\gamma^{\mu_1}\frac{i}{\slashed{\tilde p}_1-m}\gamma^{\mu_2}\frac{i}{\slashed{\tilde p}_{12}-m}\gamma^{\mu_3}\frac{i}{\slashed{\tilde p}_{123}-m}\gamma^{\mu_4},
\end{eqnarray}
\begin{eqnarray}\label{T2}
\raisebox{-0.2cm}{\incps{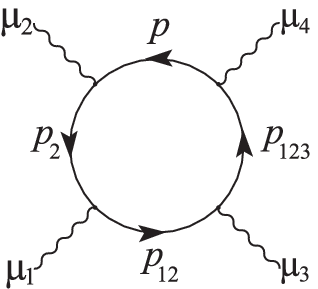}{0.0cm}{1.2cm}{3.5cm}{3.0cm}} &=& i\,T_2^{\mu_1\mu_2\mu_3\mu_4}(k_1,k_2,k_3,k_4) \nonumber\\ \nonumber\\
 &=& -(-ie)^4\int \frac{d^4p}{(2\pi)^4}{\rm tr}\frac{i}{\slashed{\tilde p}-m}\gamma^{\mu_2}\frac{i}{\slashed{\tilde p}_2-m}\gamma^{\mu_1}\frac{i}{\slashed{\tilde p}_{12}-m}\gamma^{\mu_3}\frac{i}{\slashed{\tilde p}_{123}-m}\gamma^{\mu_4},
\end{eqnarray}
\begin{eqnarray}\label{T3}
\raisebox{-0.2cm}{\incps{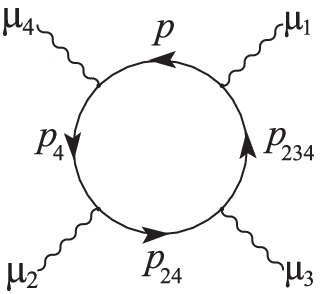}{0.0cm}{1.2cm}{3.5cm}{3.0cm}} &=& i\,T_3^{\mu_1\mu_2\mu_3\mu_4}(k_1,k_2,k_3,k_4) \nonumber\\ \nonumber\\
 &=& -(-ie)^4\int \frac{d^4p}{(2\pi)^4}{\rm tr}\frac{i}{\slashed{\tilde p}-m}\gamma^{\mu_4}\frac{i}{\slashed{\tilde p}_4-m}\gamma^{\mu_2}\frac{i}{\slashed{\tilde p}_{24}-m}\gamma^{\mu_3}\frac{i}{\slashed{\tilde p}_{234}-m}\gamma^{\mu_1}.
\end{eqnarray}
In the above expressions, $\tilde A_{\mu}(k)=((1+\kappa)A_0(k), A_i(k))$, the global factor $2$ in Eq.~(\ref{G}) means the two orientations of the fermion loop, and $p_1^\nu=p^\nu+k_1^\nu$, $p_{12}^\nu=p^\nu+k_1^\nu+k_2^\nu$, and so on.

Here, we observe that the contributions (\ref{T2}) and (\ref{T3}) can be obtained from (\ref{T1}), when we perform the interchanges:
\begin{subequations}\label{T2T3}
\begin{eqnarray}
T_2^{\mu_1\mu_2\mu_3\mu_4}(k_1,k_2,k_3,k_4) &=& T_1^{\mu_2\mu_1\mu_3\mu_4}(k_2,k_1,k_3,k_4), \\
T_3^{\mu_1\mu_2\mu_3\mu_4}(k_1,k_2,k_3,k_4) &=& T_1^{\mu_4\mu_2\mu_3\mu_1}(k_4,k_2,k_3,k_1).
\end{eqnarray}
\end{subequations}
Therefore, we just need to calculate the expression (\ref{T1}), in which, by using first the Feynman parameterization, we arrive at the expression
\begin{eqnarray}\label{T1a}
T_1^{\mu_1\mu_2\mu_3\mu_4}(k_1,k_2,k_3,k_4) &=& \int_0^1 dx_1 \int_0^{1-x_1} dx_2 \int_0^{1-x_{12}} dx_3 \int \frac{d^4p}{(2\pi)^4}\frac{6ie^4}{(\tilde p^2-M^2)^4} \nonumber\\
&&\times{\rm tr}[(\slashed{\tilde q}+m)\gamma^{\mu_1}(\slashed{\tilde q}_1+m)\gamma^{\mu_2}(\slashed{\tilde q}_{12}+m)\gamma^{\mu_3}(\slashed{\tilde q}_{123}+m)\gamma^{\mu_4}],
\end{eqnarray}
where
\begin{eqnarray}
M^2 &=& m^2-x_1(1-x_1)\tilde k_1^2-x_{12}(1-x_{12})\tilde k_2^2-x_{123}(1-x_{123})\tilde k_3^2 \nonumber\\
&&-2x_1(1-x_{12})\tilde k_1\cdot\tilde k_2-2x_1(1-x_{123})\tilde k_1\cdot\tilde k_3-2x_{12}(1-x_{123})\tilde k_2\cdot\tilde k_3,
\end{eqnarray}
and
\begin{eqnarray}
q^\mu = p^\mu-(1-x_1)k_1^\mu-(1-x_{12})k_2^\mu-(1-x_{123})k_3^\mu, \\
q_\mathrm{I}^\mu = p_\mathrm{I}^\mu-(1-x_1)k_1^\mu-(1-x_{12})k_2^\mu-(1-x_{123})k_3^\mu,
\end{eqnarray}
with $x_{12}=x_1+x_2$, $x_{123}=x_1+x_2+x_3$, and $\mathrm{I}=1,12,123$.

In order to calculate the momentum integrals in (\ref{T1a}), we will use the fact that
\begin{eqnarray}
\int \frac{d^4p}{(2\pi)^4} \frac{\tilde p_{\nu_1}\tilde p_{\nu_2}\cdots \tilde p_{\nu_p}}{(\tilde p^2-M^2)^\alpha} &=& (1+\kappa)^{-1}\int \frac{d^4\tilde p}{(2\pi)^4} \frac{\tilde p_{\nu_1}\tilde p_{\nu_2}\cdots \tilde p_{\nu_p}}{(\tilde p^2-M^2)^\alpha} \nonumber\\
&=& (1+\kappa)^{-1}\int \frac{d^4 p}{(2\pi)^4} \frac{p_{\nu_1}p_{\nu_2}\cdots p_{\nu_p}}{(p^2-M^2)^\alpha},
\end{eqnarray}
where $d^4\tilde p=d\tilde p_0dp_1dp_2dp_3$ and $d\tilde p_0=(1+\kappa)dp_0$. Consequently, the effect of Lorentz violation will be concentrated in the overall factor $(1+\kappa)^{-1}$ and into the mass parameter $M$. Thus, by calculating the trace over Dirac matrices and the integrals, we obtain, up to order $1/m^4$, the following results:
\begin{eqnarray}\label{T1b}
T_1^{\mu_1\mu_2\mu_3\mu_4}(k_1,k_2,k_3,k_4) &=& T_{1g\epsilon}^{\mu_1\mu_2\mu_3\mu_4}(k_1,k_2,k_3,k_4) + \sum_{i=1}^4T_{1gg_i}^{\mu_1\mu_2\mu_3\mu_4}(k_1,k_2,k_3,k_4) \nonumber\\
&&+ \sum_{i=1}^{12}T_{1gk_i}^{\mu_1\mu_2\mu_3\mu_4}(k_1,k_2,k_3,k_4) + \sum_{i=1}^9T_{1kk_i}^{\mu_1\mu_2\mu_3\mu_4}(k_1,k_2,k_3,k_4),
\end{eqnarray}
where
\begin{eqnarray}\label{T1ge}
T_{1g\epsilon}^{\mu_1\mu_2\mu_3\mu_4} &=& -\left[\frac{e^4}{6\pi^2\epsilon}-\frac{e^4}{12\pi^2}\ln\left(\frac{m^2}{\mu'^2}\right)\right](1+\kappa)^{-1} \nonumber\\
&&\times (g^{\mu_1\mu_2}g^{\mu_3\mu_4}-2g^{\mu_1\mu_3}g^{\mu_2\mu_4}+g^{\mu_1\mu_4}g^{\mu_2\mu_3}),
\end{eqnarray}
with $\epsilon=4-D$ and $\mu'^2=4\pi\mu^2e^{-\gamma-i\pi}$, and
\begin{subequations}\label{T1gg}
\begin{eqnarray}
T_{1gg_1}^{\mu_1\mu_2\mu_3\mu_4} &=& -\frac{e^4}{240m^2\pi^2}(1+\kappa)^{-1} \nonumber\\
&&\times[g^{\mu_1\mu_2}g^{\mu_3\mu_4}(13\tilde k_{1}^{2}+9\tilde k_{1}\cdot\tilde k_{2}+17\tilde k_{1}\cdot\tilde k_{3}+4\tilde k_{2}^{2}-\tilde k_{2}\cdot\tilde k_{3}+8\tilde k_{3}^{2}) \nonumber\\
&&+g^{\mu_1\mu_3}g^{\mu_2\mu_4} (-17\tilde k_{1}^{2}-16\tilde k_{1}\cdot\tilde k_{2}-18\tilde k_{1}\cdot\tilde k_{3}-16\tilde k_{2}^{2}-16\tilde k_{2}\cdot\tilde k_{3}-17\tilde k_{3}^{2}) \nonumber\\
&&+g^{\mu_1\mu_4}g^{\mu_2\mu_3} (8\tilde k_{1}^{2}-\tilde k_{1}\cdot\tilde k_{2}+17\tilde k_{1}\cdot\tilde k_{3}+4\tilde k_{2}^{2}+9\tilde k_{2}\cdot\tilde k_{3}+13\tilde k_{3}^{2})],
\end{eqnarray}
\begin{eqnarray}
T_{1gg_2}^{\mu_1\mu_2\mu_3\mu_4} &=& -\frac{e^4}{5040 m^4 \pi ^2}g^{{\mu_1}{\mu_2}} g^{{\mu_3}{\mu_4}}(1+\kappa)^{-1} [48 {\tilde k_1}^4+(79 {\tilde k_1}\cdot {\tilde k_2}+113 {\tilde k_1}\cdot{\tilde k_3}+48 {\tilde k_2}^2\nonumber\\
&&+13 {\tilde k_2}\cdot {\tilde k_3}+65 {\tilde k_3}^2) {\tilde k_1}^2+40 ({\tilde k_1}\cdot {\tilde k_2})^2+74 ({\tilde k_1}\cdot {\tilde k_3})^2+9
{\tilde k_2}^4-2 ({\tilde k_2}\cdot {\tilde k_3})^2\nonumber\\
&&+27 {\tilde k_3}^4+156 {\tilde k_1}\cdot {\tilde k_3} {\tilde k_2}^2+131 {\tilde k_1}\cdot {\tilde k_3} {\tilde k_2}\cdot {\tilde k_3}-3 {\tilde k_2}^2 {\tilde k_2}\cdot {\tilde k_3}+(92 {\tilde k_1}\cdot {\tilde k_3}\nonumber\\
&&+27 {\tilde k_2}^2+16 {\tilde k_2}\cdot {\tilde k_3}) {\tilde k_3}^2+{\tilde k_1}\cdot {\tilde k_2} (159 {\tilde k_1}\cdot {\tilde k_3}+39 {\tilde k_2}^2-24 {\tilde k_2}\cdot {\tilde k_3}+20
{\tilde k_3}^2)],
\end{eqnarray}
\begin{eqnarray}
T_{1gg_3}^{\mu_1\mu_2\mu_3\mu_4} &=& \frac{e^4}{5040 m^4 \pi ^2}g^{{\mu 1}{\mu 3}} g^{{\mu 2}{\mu 4}}(1+\kappa)^{-1} [57 {\tilde k_1}^4+(110 {\tilde k_1}\cdot {\tilde k_2}+118 {\tilde k_1}\cdot{\tilde k_3}+92 {\tilde k_2}^2\nonumber\\
&&+71 {\tilde k_2}\cdot {\tilde k_3}+96 {\tilde k_3}^2) {\tilde k_1}^2+72 ({\tilde k_1}\cdot {\tilde k_2})^2+80 ({\tilde k_1}\cdot {\tilde k_3})^2+54
{\tilde k_2}^4+72 ({\tilde k_2}\cdot {\tilde k_3})^2\nonumber\\
&&+57 {\tilde k_3}^4+180 {\tilde k_1}\cdot {\tilde k_3} {\tilde k_2}^2+177 {\tilde k_1}\cdot {\tilde k_3} {\tilde k_2}\cdot {\tilde k_3}+108 {\tilde k_2}^2 {\tilde k_2}\cdot {\tilde k_3}+2 (59 {\tilde k_1}\cdot {\tilde k_3}\nonumber\\
&&+46{\tilde k_2}^2+55 {\tilde k_2}\cdot {\tilde k_3}) {\tilde k_3}^2+{\tilde k_1}\cdot {\tilde k_2} (177 {\tilde k_1}\cdot {\tilde k_3}+108 {\tilde k_2}^2+66 {\tilde k_2}\cdot
{\tilde k_3}+71 {\tilde k_3}^2)],
\end{eqnarray}
\begin{eqnarray}
T_{1gg_4}^{\mu_1\mu_2\mu_3\mu_4} &=& -\frac{e^4}{5040 m^4 \pi ^2}g^{{\mu 1}{\mu 4}} g^{{\mu 2}{\mu 3}}(1+\kappa)^{-1} [27 {\tilde k_1}^4+(16 {\tilde k_1}\cdot {\tilde k_2}+92 {\tilde k_1}\cdot{\tilde k_3}+27 {\tilde k_2}^2\nonumber\\
&&+20 {\tilde k_2}\cdot {\tilde k_3}+65 {\tilde k_3}^2) {\tilde k_1}^2-2 ({\tilde k_1}\cdot {\tilde k_2})^2+74 ({\tilde k_1}\cdot {\tilde k_3})^2+9
{\tilde k_2}^4+40 ({\tilde k_2}\cdot {\tilde k_3})^2\nonumber\\
&&+48 {\tilde k_3}^4+156 {\tilde k_1}\cdot {\tilde k_3} {\tilde k_2}^2+159 {\tilde k_1}\cdot {\tilde k_3} {\tilde k_2}\cdot {\tilde k_3}+39 {\tilde k_2}^2 {\tilde k_2}\cdot {\tilde k_3}+(113 {\tilde k_1}\cdot {\tilde k_3}\nonumber\\
&&+48 {\tilde k_2}^2+79{\tilde k_2}\cdot {\tilde k_3}) {\tilde k_3}^2+{\tilde k_1}\cdot {\tilde k_2} (131 {\tilde k_1}\cdot {\tilde k_3}-3 ({\tilde k_2}^2+8 {\tilde k_2}\cdot {\tilde k_3})+13
{\tilde k_3}^2)].
\end{eqnarray}
\end{subequations}
The other contributions of Eq.~(\ref{T1b}), $T_{1gk}$ and $T_{1kk}$, have been omitted for the sake of brevity. Note that we have adopted dimensional regularization by extending the spacetime from $4$ to $D$ dimensions, so that $d^4p/(2\pi)^4$ goes to $\mu^{4-D}d^Dp/(2\pi)^D$, where $\mu$ is an arbitrary parameter that identifies the mass scale. As we will see below, from these results (\ref{T1b}) we will be able to obtain the Lorentz-violating Euler-Heisenberg action in both nonperturbative and perturbative approaches. 

Let us now recover the (perturbative in $c_{\mu\nu}$) amplitude obtained in \cite{Kostelecky:2002ue} for the photon triple splitting, from the above nonperturbative results (\ref{T1b}), in the collinear limit. For this, if we consider that the incident on-shell photon has energy $E_1$ and momentum $\vec k_1$, from the momentum conservation $\vec k_1=\vec k_2+\vec k_3+\vec k_4$, we have the inequation
\begin{equation}
|\vec k_1| = |\vec k_2+\vec k_3+\vec k_3| \le |\vec k_2|+|\vec k_3|+|\vec k_4|. 
\end{equation}
Then, as $k_i^2=E_i^2-\vec k_i^2=0$ or $|\vec k_i|=E_i$ (with $i=1,2,3,4$), in order to also satisfy the energy conservation $E_1=E_2+E_3+E_4$, all the momenta $\vec k_i$ must be aligned. Therefore, the incident photon and the decay photons must be collinear, such that the four-momenta of all photons are mutually orthogonal, $k_i^\mu k_{j\mu}=0$. By considering that these four-momenta are proportional to some four-momentum $k_0^\mu$, satisfying $k_0^2=0$, we can write $k_i^\mu=k_i k_0^\mu$, so that $k_i^\mu k_{j\mu}=k_ik_j k_0^2=0$, in which $k_i$ are now scalar coefficients, instead four-momenta. With regard to the transversality condition of the polarization four-vectors, as usual, $\epsilon_{i\mu}k_i^\mu=k_i\epsilon_{i\mu}k_0^\mu=0$ (or $A_\mu(k_i) k_i^\mu=0$). Thus, due to the requirement of collinearity, we also have that $\epsilon_{i\mu}k_j^\mu=k_j\epsilon_{i\mu}k_0^\mu=0$ (or $A_\mu(k_i) k_j^\mu=0$). 

Taking into account these considerations, we find   
\begin{eqnarray}
\nonumber G_\mathrm{coll}^{\mu_{1}\mu_{2}\mu_{3}\mu_{4}}&=&-\frac{e^4}{120m^2\pi^2}\tilde k_0^2(1+\kappa)^{-1} \nonumber\\
&&\times[g^{\mu_{1}\mu_{2}}g^{\mu_{3}\mu_{4}}(9k_{1}^{2}-8 k_{2}^{2}-k_{3}^{2}+2k_{1}k_{2}+16k_{1}k_{3}-18k_{2}k_{3}) \nonumber\\
&&+g^{\mu_{1}\mu_{3}}g^{\mu_{2}\mu_{4}}(4k_{1}^{2}-8k_{2}^{2}+4k_{3}^{2}-8k_{1}k_{2}+16k_{1}k_{3}-8k_{2}k_{3}) \nonumber\\
&&+g^{\mu_{1}\mu_{4}}g^{\mu_{2}\mu_{3}}(-k_{1}^{2}-8k_{2}^{2}+9k_{3}^{2}-18k_{1}k_{2}+16k_{1}k_{3}+2k_{2}k_{3})] \nonumber\\
&&+{\cal O}(\tilde k_0^4)+{\cal O}(\tilde k_0^{\mu_i}),
\end{eqnarray}
where we have interchanged $\mu_1$ and $\mu_2$ and thereafter $\mu_1$ and $\mu_4$, however without changing the momentum indices, in order to obtain $T_2$  and $T_3$ (Eqs.~(\ref{T2}) and (\ref{T3})) from $T_1$ (Eq.~(\ref{T1b})), following the prescription given in Ref.~\cite{Kostelecky:2002ue}. Note that, as expected, the divergent term (\ref{T1ge}) vanishes, and, for simplicity, we have not included the terms of ${\cal O}(\tilde k_0^4)$ and ${\cal O}(\tilde k_0^{\mu_i})$. Now, when we expand up to first order in $\kappa$, the Lorentz-violating contributions become
\begin{eqnarray}
\tilde k_0^2(1+\kappa)^{-1}&=&[(k_0^0)^2(1+\kappa)^2-(k_0^i)^2](1+\kappa)^{-1} \nonumber\\
&=& 2\kappa(k_0^0)^2+{\cal O}(\kappa^2).
\end{eqnarray}
Therefore, the perturbative result is readily recovered (see Eq.~(10) of Ref.~\cite{Kostelecky:2002ue}), which is given by
\begin{eqnarray}\label{Gcoll}
\nonumber G_{\mathrm{coll},\kappa}^{\mu_{1}\mu_{2}\mu_{3}\mu_{4}}&=&-\frac{e^4c_{\mu\nu}k_0^\mu k_0^\nu}{60\pi^2m^2}[g^{\mu_{1}\mu_{2}}g^{\mu_{3}\mu_{4}}(9k_{1}^{2}-8 k_{2}^{2}-k_{3}^{2}+2k_{1}k_{2}+16k_{1}k_{3}-18k_{2}k_{3}) \nonumber\\
&&+g^{\mu_{1}\mu_{3}}g^{\mu_{2}\mu_{4}}(4k_{1}^{2}-8k_{2}^{2}+4k_{3}^{2}-8k_{1}k_{2}+16k_{1}k_{3}-8k_{2}k_{3}) \nonumber\\
&&+g^{\mu_{1}\mu_{4}}g^{\mu_{2}\mu_{3}}(-k_{1}^{2}-8k_{2}^{2}+9k_{3}^{2}-18k_{1}k_{2}+16k_{1}k_{3}+2k_{2}k_{3})],
\end{eqnarray}
with $c_{\mu\nu}k_0^\mu k_0^\nu=\kappa(k_0^0)^2$, however, for $c_{00}\neq0$ and $c_{0i}=c_{ij}=0$.

We will now discuss the generation of the Lorentz-violating Euler-Heisenberg action, in the first order correction, $\alpha^2=e^4/16\pi^2$. For this, we must calculate $G$ (Eq.~(\ref{G})) without regard to the collinear limit. Thus, as pointed out at the beginning, $T_2$ and $T_3$ are obtained from $T_1$, when we interchange the uncontracted indices as well as the momentum indices, see Eq.~(\ref{T2T3}). Proceeding in this way, we obtain
\begin{eqnarray}\label{Ga}
G^{\mu_1\mu_2\mu_3\mu_4}(k_1,k_2,k_3,k_4) &=& G_{gg}^{\mu_1\mu_2\mu_3\mu_4}(k_1,k_2,k_3,k_4) + \sum_{i=1}^6G_{gk_i}^{\mu_1\mu_2\mu_3\mu_4}(k_1,k_2,k_3,k_4) \nonumber\\
&&+ G_{kk}^{\mu_1\mu_2\mu_3\mu_4}(k_1,k_2,k_3,k_4),
\end{eqnarray}
where
\begin{eqnarray}
G_{gg}^{\mu_1\mu_2\mu_3\mu_4} &=& \frac{e^4}{180\pi^2m^4}(1+\kappa)^{-1} \nonumber\\
&&\times[g^{\mu_1\mu_2}g^{\mu_3\mu_4}(7\tilde k_1\cdot \tilde k_4 \tilde k_2\cdot \tilde k_3+7\tilde k_1\cdot \tilde k_3 \tilde k_2\cdot \tilde k_4-10\tilde k_1\cdot \tilde k_2 \tilde k_3\cdot \tilde k_4) \nonumber\\
&&+g^{\mu_1\mu_3}g^{\mu_2\mu_4}(7\tilde k_1\cdot \tilde k_4 \tilde k_2\cdot \tilde k_3-10\tilde k_1\cdot \tilde k_3 \tilde k_2\cdot \tilde k_4+7\tilde k_1\cdot \tilde k_2 \tilde k_3\cdot \tilde k_4) \nonumber\\
&&+g^{\mu_1\mu_4}g^{\mu_2\mu_3}(-10\tilde k_1\cdot \tilde k_4 \tilde k_2\cdot \tilde k_3+7\tilde k_1\cdot \tilde k_3 \tilde k_2\cdot \tilde k_4+7\tilde k_1\cdot \tilde k_2 \tilde k_3\cdot \tilde k_4)],
\end{eqnarray}
\begin{subequations}\label{Ggk}
\begin{eqnarray}\label{Ggk1}
G_{gk_1}^{\mu_1\mu_2\mu_3\mu_4} &=& \frac{e^4}{180\pi^2m^4}g^{\mu_{1}\mu_{2}}(1+\kappa)^{-1} [\tilde k_4^{\mu_3}(10\tilde k_3^{\mu_4}\tilde k_1\cdot \tilde k_2-7(\tilde k_2^{\mu_4}\tilde k_1\cdot \tilde k_3+\tilde k_1^{\mu_4}\tilde k_2\cdot \tilde k_3))\nonumber\\
&&-7\tilde k_3^{\mu_4}(\tilde k_2^{\mu_3}\tilde k_1\cdot \tilde k_4+\tilde k_1^{\mu_3}\tilde k_2\cdot \tilde k_4)+7(\tilde k_2^{\mu_3}\tilde k_1^{\mu_4}+\tilde k_1^{\mu_3}\tilde k_2^{\mu_4})\tilde k_3\cdot \tilde k_4],
\end{eqnarray}
\begin{eqnarray}\label{Ggk2}
G_{gk_2}^{\mu_1\mu_2\mu_3\mu_4} &=& \frac{e^4}{180\pi^2m^4}g^{\mu_{1}\mu_{3}}(1+\kappa)^{-1} [\tilde k_4^{\mu_2}(-7\tilde k_3^{\mu_4}\tilde k_1\cdot \tilde k_2+10\tilde k_2^{\mu_4}\tilde k_1\cdot \tilde k_3-7\tilde k_1^{\mu_4}\tilde k_2\cdot \tilde k_3)\nonumber\\
&&+\tilde k_3^{\mu_2}(7\tilde k_1^{\mu_4}\tilde k_2\cdot \tilde k_4-7\tilde k_2^{\mu_4}\tilde k_1\cdot \tilde k_4)+7\tilde k_1^{\mu_2}(\tilde k_3^{\mu_4}\tilde k_2\cdot \tilde k_4-\tilde k_2^{\mu_4}\tilde k_3\cdot \tilde k_4)],
\end{eqnarray}
\begin{eqnarray}\label{Ggk3}
G_{gk_3}^{\mu_1\mu_2\mu_3\mu_4} &=& \frac{e^4}{180\pi^2m^4}g^{\mu_{1}\mu_{4}}(1+\kappa)^{-1} [\tilde k_4^{\mu_2}(7\tilde k_1^{\mu_3}\tilde k_2\cdot \tilde k_3-7\tilde k_2^{\mu_3}\tilde k_1\cdot \tilde k_3)+\tilde k_3^{\mu_2}(-7\tilde k_4^{\mu_3}\tilde k_1\cdot \tilde k_2\nonumber\\
&&+10\tilde k_2^{\mu_3}\tilde k_1\cdot \tilde k_4-7\tilde k_1^{\mu_3}\tilde k_2\cdot \tilde k_4)+7\tilde k_1^{\mu_2}(\tilde k_4^{\mu_3}\tilde k_2\cdot \tilde k_3-\tilde k_2^{\mu_3}\tilde k_3\cdot \tilde k_4)],
\end{eqnarray}
\begin{eqnarray}\label{Ggk4}
G_{gk_4}^{\mu_1\mu_2\mu_3\mu_4} &=& \frac{e^4}{180\pi^2m^4}g^{\mu_{2}\mu_{3}}(1+\kappa)^{-1} [\tilde k_4^{\mu_1}(-7\tilde k_3^{\mu_4}\tilde k_1\cdot \tilde k_2-7\tilde k_2^{\mu_4}\tilde k_1\cdot \tilde k_3+10\tilde k_1^{\mu_4}\tilde k_2\cdot \tilde k_3)\nonumber\\
&&+7(\tilde k_3^{\mu_1}(\tilde k_2^{\mu_4}\tilde k_1\cdot \tilde k_4-\tilde k_1^{\mu_4}\tilde k_2\cdot \tilde k_4)+\tilde k_2^{\mu_1}(\tilde k_3^{\mu_4}\tilde k_1\cdot \tilde k_4-\tilde k_1^{\mu_4}\tilde k_3\cdot \tilde k_4))],
\end{eqnarray}
\begin{eqnarray}\label{Ggk5}
G_{gk_5}^{\mu_1\mu_2\mu_3\mu_4} &=& \frac{e^4}{180\pi^2m^4}g^{\mu_{2}\mu_{4}}(1+\kappa)^{-1} [-7\tilde k_4^{\mu_1}(\tilde k_1^{\mu_3}\tilde k_2\cdot \tilde k_3-\tilde k_2^{\mu_3}\tilde k_1\cdot \tilde k_3)-\tilde k_3^{\mu_1}(7\tilde k_4^{\mu_3}\tilde k_1\cdot \tilde k_2\nonumber\\
&&+7\tilde k_2^{\mu_3}\tilde k_1\cdot \tilde k_4-10\tilde k_1^{\mu_3}\tilde k_2\cdot \tilde k_4)-7\tilde k_2^{\mu_1}(\tilde k_1^{\mu_3}\tilde k_3\cdot \tilde k_4-\tilde k_4^{\mu_3}\tilde k_1\cdot \tilde k_3)],
\end{eqnarray}
\begin{eqnarray}\label{Ggk6}
G_{gk_6}^{\mu_1\mu_2\mu_3\mu_4} &=& \frac{e^4}{180\pi^2m^4}g^{\mu_{3}\mu_{4}}(1+\kappa)^{-1} [7\tilde k_4^{\mu_1}(\tilde k_3^{\mu_2}\tilde k_1\cdot \tilde k_2-\tilde k_1^{\mu_2}\tilde k_2\cdot \tilde k_3)+7\tilde k_3^{\mu_1}(\tilde k_4^{\mu_2}\tilde k_1\cdot \tilde k_2\nonumber\\
&&-\tilde k_1^{\mu_2}\tilde k_2\cdot \tilde k_4)+\tilde k_2^{\mu_1}(-7\tilde k_4^{\mu_2}\tilde k_1\cdot \tilde k_3-7\tilde k_3^{\mu_2}\tilde k_1\cdot \tilde k_4+10\tilde k_1^{\mu_2}\tilde k_3\cdot \tilde k_4)],
\end{eqnarray}
\end{subequations}
\begin{eqnarray}\label{T1kk}
G_{kk}^{\mu_1\mu_2\mu_3\mu_4} &=& \frac{e^4}{180\pi^2m^4}(1+\kappa)^{-1}\nonumber\\
&&\times[\tilde k_2^{\mu_1}(7\tilde k_3^{\mu_2}\tilde k_4^{\mu_3}\tilde k_1^{\mu_4}+(7\tilde k_4^{\mu_2}\tilde k_1^{\mu_3}-10\tilde k_1^{\mu_2}\tilde k_4^{\mu_3})\tilde k_3^{\mu_4})\nonumber\\
&&+\tilde k_3^{\mu_1}(7\tilde k_1^{\mu_2}\tilde k_4^{\mu_3}\tilde k_2^{\mu_4}+\tilde k_4^{\mu_2}(7\tilde k_2^{\mu_3}\tilde k_1^{\mu_4}-10\tilde k_1^{\mu_3}\tilde k_2^{\mu_4}))\nonumber\\
&&+\tilde k_4^{\mu_1}(\tilde k_3^{\mu_2}(7\tilde k_1^{\mu_3}\tilde k_2^{\mu_4}-10\tilde k_2^{\mu_3}\tilde k_1^{\mu_4})+7\tilde k_1^{\mu_2}\tilde k_2^{\mu_3}\tilde k_3^{\mu_4})],
\end{eqnarray}
with $\tilde k_4=-\tilde k_1-\tilde k_2-\tilde k_3$. We observe that all the contributions of order $1/m^2$ completely vanish, as expected. In fact, the Euler-Heisenberg actions are all proportional to order $1/m^4$, in the first order correction.

Therefore, by considering these results (\ref{Ga}), the effective action (\ref{Seff}) takes the following form
\begin{equation}\label{SEH}
S_{\rm EH} = -\frac{\alpha^2}{180m^4}(1+\kappa)^{-1} \int d^4x \int d^4k_1d^4k_2d^4k_3d^4k_4 \; e^{i(k_1+k_2+k_3+k_4)\cdot x}  G(k_1,k_2,k_3,k_4) ,
\end{equation}
where
\begin{equation}
G(k_1,k_2,k_3,k_4)=G^{\mu_1\mu_2\mu_3\mu_4}(k_1,k_2,k_3,k_4) \tilde A_{\mu_1}(k_1)\tilde A_{\mu_2}(k_2)\tilde A_{\mu_3}(k_3)\tilde A_{\mu_4}(k_4), 
\end{equation}
so that we can write
\begin{eqnarray}
G(k_1,k_2,k_3,k_4) &=& \frac53 \tilde F_{\mu\nu}(k_1)\tilde F^{\mu\nu}(k_2) \tilde F_{\lambda\rho}(k_3)\tilde F^{\lambda\rho}(k_4) + \frac53 \tilde F_{\mu\nu}(k_1)\tilde F^{\mu\nu}(k_3) \tilde F_{\lambda\rho}(k_2)\tilde F^{\lambda\rho}(k_4) \nonumber\\
&&+ \frac53 \tilde F_{\mu\nu}(k_1)\tilde F^{\mu\nu}(k_4) \tilde F_{\lambda\rho}(k_2)\tilde F^{\lambda\rho}(k_3) - \frac{14}6 F_{\mu\nu}(k_1)\tilde F^{\nu\lambda}(k_2) \tilde F_{\lambda\rho}(k_3)\tilde F^{\rho\mu}(k_4) \nonumber\\
&&- \frac{14}6 F_{\mu\nu}(k_1)\tilde F^{\nu\lambda}(k_2) \tilde F_{\lambda\rho}(k_4)\tilde F^{\rho\mu}(k_3) - \frac{14}6 F_{\mu\nu}(k_1)\tilde F^{\nu\lambda}(k_3) \tilde F_{\lambda\rho}(k_2)\tilde F^{\rho\mu}(k_4) \nonumber\\
&&- \frac{14}6 F_{\mu\nu}(k_1)\tilde F^{\nu\lambda}(k_3) \tilde F_{\lambda\rho}(k_4)\tilde F^{\rho\mu}(k_2) - \frac{14}6 F_{\mu\nu}(k_1)\tilde F^{\nu\lambda}(k_4) \tilde F_{\lambda\rho}(k_2)\tilde F^{\rho\mu}(k_3) \nonumber\\
&&- \frac{14}6 F_{\mu\nu}(k_1)\tilde F^{\nu\lambda}(k_4) \tilde F_{\lambda\rho}(k_3)\tilde F^{\rho\mu}(k_2),
\end{eqnarray}
with $\tilde F^{\mu\nu}(k_1)=\tilde k_1^\mu \tilde A^\nu(k_1)-\tilde k_1^\nu \tilde A^\mu(k_1)$, and so on. Then, inverting the Fourier transform in the expression (\ref{SEH}), the Lorentz-violating Euler-Heisenberg action becomes
\begin{equation}\label{SEH2}
S_{\rm EH} = -\frac{\alpha^2}{180m^4}(1+\kappa)^{-1} \int d^4x (5\tilde F_{\mu\nu}\tilde F^{\mu\nu} \tilde F_{\lambda\rho}\tilde F^{\lambda\rho} - 14\tilde F_{\mu\nu}\tilde F^{\nu\lambda} \tilde F_{\lambda\rho}\tilde F^{\rho\mu}),
\end{equation}
where now $\tilde F^{\mu\nu}=\tilde\partial^\mu \tilde A^\nu(x)-\tilde\partial^\nu \tilde A^\mu(x)$. The above result is somewhat unusual, although it resembles the conventional Euler-Heisenberg action \cite{Karplus:1950zza}. From this expression (\ref{SEH2}) (or from the equations (\ref{T1b})) we are now able to calculate some scattering amplitudes of interest in the presence of Lorentz violation (see Sec.~\ref{intro}, for some proposals).

It is worth mentioning that the Lagrangian of Eq.~(\ref{SEH2}) is the leading term of the expansion of the Euler-Heisenberg effective Lagrangian, in the weak-field limit, which, written in terms of the dual field strength ${^*\tilde F_{\mu\nu}}=\frac12\epsilon_{\mu\nu\lambda\rho}\tilde F^{\lambda\rho}$ and the critical field strength $E_c=m^2c^3/\hbar e=1.3\times10^{16}$~V/cm, can be reexpressed as
\begin{equation}\label{LEH2}
{\cal L}_{\rm EH}^{(2)} = \frac{\varkappa}{16}(1+\kappa)^{-1}(4\tilde F_{\mu\nu}\tilde F^{\mu\nu} \tilde F_{\lambda\rho}\tilde F^{\lambda\rho} + 7\tilde F_{\mu\nu}{^*\tilde F^{\mu\nu}} \tilde F_{\lambda\rho}{^*\tilde F^{\lambda\rho}}),
\end{equation}
where
\begin{equation}
\varkappa=\frac{2\alpha^2\hbar^3}{45m^4c^5} = \frac{\alpha}{90\pi E_c^2},
\end{equation}
with $\alpha=e^2/4\pi\hbar c$ (note that we have reinstated $\hbar$ and $c$). In this limit, the electric and magnetic fields $E$ and $B$ are weak in comparison with the critical field $E_c$. In fact, the limit $E\ll E_c$ is essential for the validity of the effective Lagrangian as well as imposes severe restrictions on the real (electron-positron) pair production. We refer the reader to Ref.~\cite{Berestetskii}, for a more complete discussion. 

While the above expression~(\ref{LEH2}) is sufficient to analyze the processes of (four-photon) box diagrams, for more external legs, we can infer from the Euler-Heisenberg effective Lagrangian the other expressions, e.g., for (six-photon) hexagon diagrams, we can write
\begin{equation}\label{LEH3}
{\cal L}_{\rm EH}^{(3)} = -\frac{\xi}{64}(1+\kappa)^{-1}\tilde F_{\alpha\beta}\tilde F^{\alpha\beta} (8\tilde F_{\mu\nu}\tilde F^{\mu\nu} \tilde F_{\lambda\rho}\tilde F^{\lambda\rho} + 13\tilde F_{\mu\nu}{^*\tilde F^{\mu\nu}} \tilde F_{\lambda\rho}{^*\tilde F^{\lambda\rho}}),
\end{equation}
where
\begin{equation}
\xi=\frac{32\pi\alpha^3\hbar^6}{315m^8c^{10}} = \frac{2\alpha}{315\pi E_c^4}.
\end{equation}
In particular, the Lagrangiana (\ref{LEH3}) is that responsible for the first non-zero contribution to the photon splitting in an (constant and spatially uniform) external magnetic field $\bar B$~\cite{BialynickaBirula:1970vy,Adler:1970gg}, in which $\bar B\sim B_c=m^2c^3/\hbar e=4.42\times10^{13}$~G (i.e., a strong field). The limit of very strong magnetic fields ($B\gg B_c$) is also interesting, however, it must be taken into account in the complete Euler-Heisenberg effective Lagrangian. We believe that the Lorentz-violating version of this effective Lagrangian resembles its original shape, except for the overall factor $(1+\kappa)^{-1}$ and for the field strength $\tilde F^{\mu\nu}=((1+\kappa)F^{0i},F^{ij})$ and its dual. Nevertheless, this will be confirmed in a forthcoming study.

In general, calculations involving Lorentz violation are performed in the first order of the coefficient that controls the violation. Thus, by expanding the action~(\ref{SEH2}) up to first order in $\kappa$, we obtain 
\begin{eqnarray}\label{SEH3}
S_{\rm EH} &=& -\frac{\alpha^2}{180m^4} \int d^4x\,(5F_{\mu\nu}F^{\mu\nu}F_{\lambda\rho}F^{\lambda\rho} - 14F_{\mu\nu}F^{\nu\lambda}F_{\lambda\rho}F^{\rho\mu}) \nonumber\\
&&-\frac{\alpha^2}{90m^4} \int d^4x\, (k_F)_{\mu\nu\alpha\beta}(5F^{\alpha\beta}F^{\mu\nu}F_{\lambda\rho}F^{\lambda\rho} - 14 F^{\alpha\beta}F^{\nu\lambda}F_{\lambda\rho}F^{\rho\mu}) \nonumber\\
&&+\frac{\alpha^2}{180m^4} \int d^4x\, c_{\alpha\beta}g^{\alpha\beta}(5F_{\mu\nu}F^{\mu\nu}F_{\lambda\rho}F^{\lambda\rho} - 14F_{\mu\nu}F^{\nu\lambda}F_{\lambda\rho}F^{\rho\mu}) + {\cal O}(c_{\mu\nu}^2),
\end{eqnarray}
where we have introduced the coefficient $(k_F)_{\mu\nu\alpha\beta}$, given by
\begin{equation}\label{kF}
(k_F)_{\mu\nu\alpha\beta} = g_{\mu\alpha}c_{\nu\beta}+g_{\nu\beta}c_{\mu\alpha}-g_{\mu\beta}c_{\nu\alpha}-g_{\nu\alpha}c_{\mu\beta}.
\end{equation}
 
In the next section we will see that the second and third contributions of Eq.~(\ref{SEH3}) are in fact found from the perturbative approach, however, for a generic coefficient $c_{\mu\nu}$. An interesting point about the decomposition of~$(k_F)_{\mu\nu\alpha\beta}$ (Eq.~(\ref{kF})), in terms of a symmetry tensor $c_{\mu\nu}$, is that it is also found in the Born-Infeld electrodynamics, with an (Lorentz-violating) external field \cite{VillalbaChavez:2012ea}. In particular, this decomposition restricts the Lorentz-violating electrodynamics to a nonbirefringence sector \cite{Klinkhamer:2008ky}.

\section{Perturbative approach}\label{perturbative}

In order to take into account the perturbative approach, let us first rewrite the fermion Lagrangian (\ref{L}) as follows
\begin{equation}\label{L2}
{\cal L}_f=\bar\psi(i\partial_\mu\gamma^\mu + ic_{\mu\nu}\partial^\mu\gamma^\nu- m - e A_\mu \gamma^\mu - e A_\mu c^{\mu\nu}\gamma_\nu)\psi,
\end{equation}
in which, from now on, $c_{\mu\nu}$ is a generic coefficient for Lorentz violation. Now, by expanding the propagator up to first order in $c_{\mu\nu}$,
\begin{equation}
\frac{i}{\slashed{p}+c_{\mu\nu}p^\mu\gamma^\nu-m} = \frac{i}{\slashed{p}-m}+\frac{i}{\slashed{p}-m}ic_{\mu\nu}p^\mu\gamma^\nu\frac{i}{\slashed{p}-m} + \cdots,
\end{equation}
so that we can consider $ic_{\mu\nu}p^\mu\gamma^\nu$ as a insertion in the propagator $iS(p)=i(\slashed{p}-m)^{-1}$, and treating  the last term in~(\ref{L2}) as a new vertex, the effective action (\ref{Seff}) then becomes
\begin{eqnarray}
S_{\rm eff}^{(4)} &=& \frac14 \int d^4x \int d^4k_1d^4k_2d^4k_3d^4k_4 \; e^{i(k_1+k_2+k_3+k_4)\cdot x} \nonumber\\
&&\times \frac16 G_c^{\mu_1\mu_2\mu_3\mu_4}(k_1,k_2,k_3,k_4) A_{\mu_1}(k_1) A_{\mu_2}(k_2) A_{\mu_3}(k_3) A_{\mu_4}(k_4) + {\cal O}\left(c_{\mu\nu}^2\right)
\end{eqnarray}
with
\begin{eqnarray}\label{Gc}
G_c^{\mu_1\mu_2\mu_3\mu_4}(k_1,k_2,k_3,k_4) &=& 2\,T_{c1}^{\mu_1\mu_2\mu_3\mu_4}(k_1,k_2,k_3,k_4) +  2\,T_{c2}^{\mu_1\mu_2\mu_3\mu_4}(k_1,k_2,k_3,k_4) \nonumber\\
&&+ 2\,T_{c3}^{\mu_1\mu_2\mu_3\mu_4}(k_1,k_2,k_3,k_4),
\end{eqnarray}
where now each graph above is subdivided into eight graphs,
\begin{eqnarray}
T_{c1}^{\mu_1\mu_2\mu_3\mu_4}(k_1,k_2,k_3,k_4)&=&T_{c11}^{\mu_1\mu_2\mu_3\mu_4}(k_1,k_2,k_3,k_4)+T_{c12}^{\mu_1\mu_2\mu_3\mu_4}(k_1,k_2,k_3,k_4) \nonumber\\
&&+T_{c13}^{\mu_1\mu_2\mu_3\mu_4}(k_1,k_2,k_3,k_4)+T_{c14}^{\mu_1\mu_2\mu_3\mu_4}(k_1,k_2,k_3,k_4),
\end{eqnarray}
given by
\begin{eqnarray}\label{Tc11}
T_{c11}^{\mu_1\mu_2\mu_3\mu_4} &=& ie^4\int \frac{d^4p}{(2\pi)^4}{\rm tr}\,S(p)c^{\mu_1\mu}\gamma_\mu S(p_1)\gamma^{\mu_2} S(p_{12})\gamma^{\mu_3} S(p_{123})\gamma^{\mu_4} \nonumber\\
&&-ie^4\int \frac{d^4p}{(2\pi)^4}{\rm tr}\,S(p)c_{\mu\nu}p^\mu\gamma^\nu S(p)\gamma^{\mu_1} S(p_1)\gamma^{\mu_2} S(p_{12})\gamma^{\mu_3} S(p_{123})\gamma^{\mu_4},
\end{eqnarray}
\begin{eqnarray}
T_{c12}^{\mu_1\mu_2\mu_3\mu_4} &=& ie^4\int \frac{d^4p}{(2\pi)^4}{\rm tr}\,S(p)\gamma^{\mu_1} S(p_1)c^{\mu_2\mu}\gamma_\mu S(p_{12})\gamma^{\mu_3} S(p_{123})\gamma^{\mu_4} \nonumber\\
&&-ie^4\int \frac{d^4p}{(2\pi)^4}{\rm tr}\,S(p)\gamma^{\mu_1} S(p_1)c_{\mu\nu}p_1^\mu\gamma^\nu S(p_1)\gamma^{\mu_2} S(p_{12})\gamma^{\mu_3} S(p_{123})\gamma^{\mu_4},
\end{eqnarray}
\begin{eqnarray}
T_{c13}^{\mu_1\mu_2\mu_3\mu_4} &=& ie^4\int \frac{d^4p}{(2\pi)^4}{\rm tr}\,S(p)\gamma^{\mu_1} S(p_1)\gamma^{\mu_2} S(p_{12})c^{\mu_3\mu}\gamma_\mu S(p_{123})\gamma^{\mu_4} \nonumber\\
&&-ie^4\int \frac{d^4p}{(2\pi)^4}{\rm tr}\,S(p)\gamma^{\mu_1} S(p_1)\gamma^{\mu_2} S(p_{12})c_{\mu\nu}p_{12}^\mu\gamma^\nu S(p_{12})\gamma^{\mu_3} S(p_{123})\gamma^{\mu_4},
\end{eqnarray}
\begin{eqnarray}
T_{c14}^{\mu_1\mu_2\mu_3\mu_4} &=& ie^4\int \frac{d^4p}{(2\pi)^4}{\rm tr}\,S(p)\gamma^{\mu_1} S(p_1)\gamma^{\mu_2} S(p_{12})\gamma^{\mu_3} S(p_{123})c^{\mu_4\mu}\gamma_\mu \nonumber\\
&&-ie^4\int \frac{d^4p}{(2\pi)^4}{\rm tr}\,S(p)\gamma^{\mu_1} S(p_1)\gamma^{\mu_2} S(p_{12})\gamma^{\mu_3} S(p_{123})c_{\mu\nu}p_{123}^\mu\gamma^\nu S(p_{123})\gamma^{\mu_4}.
\end{eqnarray}
Once more, it is easy to see that  $T_{c2}$ and $T_{c3}$ are obtained from $T_{c1}$, as well as $T_{c12}$, $T_{c13}$, and $T_{c14}$ from $T_{c11}$, when we also perform the ciclic interchanges: 
\begin{subequations}\label{Tc1234}
\begin{eqnarray}
T_{c12}^{\mu_1\mu_2\mu_3\mu_4}(k_1,k_2,k_3,k_4)&=&T_{c11}^{\mu_4\mu_1\mu_2\mu_3}(k_4,k_1,k_2,k_3),\\
T_{c13}^{\mu_1\mu_2\mu_3\mu_4}(k_1,k_2,k_3,k_4)&=&T_{c11}^{\mu_3\mu_4\mu_1\mu_2}(k_3,k_4,k_1,k_2),\\
T_{c14}^{\mu_1\mu_2\mu_3\mu_4}(k_1,k_2,k_3,k_4)&=&T_{c11}^{\mu_2\mu_3\mu_4\mu_1}(k_2,k_3,k_4,k_1).
\end{eqnarray}
\end{subequations}

Therefore, we must focus our attention on the two graphs of $T_{c11}$ (Eq.~(\ref{Tc11})), in which, by considering first the Feynman parameterization, we obtain
\begin{eqnarray}
T_{c11}^{\mu_1\mu_2\mu_3\mu_4} &=& \int_0^1 dx_1 \int_0^{1-x_1} dx_2 \int_0^{1-x_{12}} dx_3 \int \frac{d^4p}{(2\pi)^4}\frac{6ie^4}{(p^2-M^2)^4}\nonumber\\
&&\times{\rm tr}[(\slashed{q}+m)c^{\mu_1\mu}\gamma_\mu(\slashed{q}_1+m)\gamma^{\mu_2}(\slashed{q}_{12}+m)\gamma^{\mu_3}(\slashed{q}_{123}+m)\gamma^{\mu_4}]\nonumber\\
&&- \int_0^1 dx_1 \int_0^{1-x_1} dx_2 \int_0^{1-x_{12}} dx_3 \int \frac{d^4p}{(2\pi)^4}\frac{24ixe^4}{(p^2-M^2)^5}\nonumber\\
&&\times{\rm tr}[(\slashed{q}+m)c_{\mu\nu}q^\mu\gamma^\nu(\slashed{q}+m)\gamma^{\mu_1}(\slashed{q}_1+m)\gamma^{\mu_2}(\slashed{q}_{12}+m)\gamma^{\mu_3}(\slashed{q}_{123}+m)\gamma^{\mu_4}].
\end{eqnarray}
Following the standard procedure, after we calculate the trace over Dirac matrices and the corresponding integrals, up to order $1/m^4$, we arrive at
\begin{eqnarray}
T_{c11}^{\mu_1\mu_2\mu_3\mu_4}(k_1,k_2,k_3,k_4) &=& T_{c11g\epsilon}^{\mu_1\mu_2\mu_3\mu_4}(k_1,k_2,k_3,k_4) + \sum_{i=1}^8T_{c11gg_i}^{\mu_1\mu_2\mu_3\mu_4}(k_1,k_2,k_3,k_4) \nonumber\\
&&+ \sum_{i=1}^{24}T_{c11gk_i}^{\mu_1\mu_2\mu_3\mu_4}(k_1,k_2,k_3,k_4) + \sum_{i=1}^{18}T_{c11kk_i}^{\mu_1\mu_2\mu_3\mu_4}(k_1,k_2,k_3,k_4),
\end{eqnarray}
where
\begin{eqnarray}\label{T1age}
T_{c11g\epsilon}^{\mu_1\mu_2\mu_3\mu_4} &=& -\left[\frac{e^4}{12\pi^2\epsilon}-\frac{e^4}{24\pi^2}\ln\left(\frac{m^2}{\mu'^2}\right)\right](-g^{\mu _1\mu _2} c^{\mu _3\mu _4}+c^{\mu _1\mu _2} g^{\mu _3\mu _4}-4 c^{\mu_1\mu _3}g^{\mu _2\mu _4}\nonumber\\
&&+2 c^{\mu_1\mu_4}g^{\mu _2\mu _3}+2g^{\mu _1\mu _4} c^{\mu _2\mu _3}+c_{\mu\nu}g^{\mu\nu} g^{\mu _1\mu _3} g^{\mu _2\mu _4}-c_{\mu\nu}g^{\mu\nu} g^{\mu _1\mu _4}g^{\mu _2\mu _3}),
\end{eqnarray}
\begin{subequations}\label{c11gg}
\begin{eqnarray}
T_{c11gg_1}^{\mu_1\mu_2\mu_3\mu_4} &=& \frac{e^4c_{\mu\nu}g^{\mu\nu}}{1440 m^2 \pi ^2} [g^{{\mu_1}{\mu_2}} g^{{\mu_3}{\mu_4}}(27 k_1^2+17 k_1\cdot k_2+37 k_1\cdot k_3+9 k_2\cdot k_3+10 k_3^2) \nonumber\\
&&+g^{{\mu_1}{\mu_3}} g^{{\mu_2}{\mu_4}}(-39 k_1^2-37 k_1\cdot k_2-41 k_1\cdot k_3-24 k_2^2-39 k_2\cdot k_3-26 k_3^2) \nonumber\\
&&+g^{{\mu_1}{\mu_4}} g^{{\mu_2}{\mu_3}}(24 k_1^2+11 k_1\cdot k_2+37 k_1\cdot k_3+12 k_2^2+33 k_2\cdot k_3+25 k_3^2)],
\end{eqnarray}
\begin{eqnarray}
T_{c11gg_2}^{\mu_1\mu_2\mu_3\mu_4} &=& \frac{e^4c_{\mu\nu}g^{\mu\nu}}{10080 m^4 \pi ^2}g^{{\mu_1}{\mu_2}} g^{{\mu_3}{\mu_4}} [39 k_1^4+(63 k_1\cdot
k_2+93 k_1\cdot k_3+30 k_2^2+22 k_2\cdot k_3\nonumber\\
&&+45 k_3^2) k_1^2+31 (k_1\cdot k_2)^2+61 (k_1\cdot k_3)^2+15 (k_2\cdot k_3)^2+14 k_3^4+93 k_1\cdot k_3 k_2^2\nonumber\\
&&+96 k_1\cdot k_3 k_2\cdot k_3+12 k_2^2 k_2\cdot k_3+2(33 k_1\cdot k_3+7 k_2^2+15 k_2\cdot k_3) k_3^2\nonumber\\
&&+k_1\cdot k_2 (127 k_1\cdot k_3+23 k_2^2+11 k_2\cdot k_3+31 k_3^2)],
\end{eqnarray}
\begin{eqnarray}
T_{c11gg_3}^{\mu_1\mu_2\mu_3\mu_4} &=& -\frac{e^4c_{\mu\nu}g^{\mu\nu}}{10080 m^4 \pi ^2}g^{\mu_1\mu_3} g^{\mu_2\mu_4}[48 k_1^4+(93 k_1\cdot k_2+99 k_1\cdot k_3+65 (k_2^2+k_2\cdot k_3)\nonumber\\
&&+68 k_3^2) k_1^2+61 (k_1\cdot k_2)^2+67 (k_1\cdot k_3)^2+27 k_2^4+57 (k_2\cdot k_3)^2+29 k_3^4\nonumber\\
&&+113 k_1\cdot k_3 k_2^2+132 k_1\cdot k_3 k_2\cdot k_3+75 k_2^2 k_2\cdot k_3+2 (43 k_1\cdot k_3+25 k_2^2\nonumber\\
&&+40 k_2\cdot k_3) k_3^2+k_1\cdot k_2 (145 k_1\cdot k_3+79 (k_2^2+k_2\cdot k_3)+69 k_3^2)],
\end{eqnarray}
\begin{eqnarray}
T_{c11gg_4}^{\mu_1\mu_2\mu_3\mu_4} &=& \frac{e^4c_{\mu\nu}g^{\mu\nu}}{10080 m^4 \pi ^2}g^{\mu_1\mu_4} g^{\mu_2\mu_3}[27 k_1^4+(29 k_1\cdot k_2+79 k_1\cdot k_3+27 k_2^2+35 k_2\cdot k_3\nonumber\\
&&+52 k_3^2)k_1^2+11 (k_1\cdot k_2)^2+61 (k_1\cdot k_3)^2+9 k_2^4+45 (k_2\cdot k_3)^2+28 k_3^4\nonumber\\
&&+95 k_1\cdot k_3 k_2^2+116 k_1\cdot k_3 k_2\cdot k_3+48 k_2^2 k_2\cdot k_3+(80 k_1\cdot k_3+36 k_2^2\nonumber\\
&&+71 k_2\cdot k_3) k_3^2+k_1\cdot k_2 (111 k_1\cdot k_3+17 k_2^2+29 k_2\cdot k_3+41 k_3^2)].
\end{eqnarray}
\end{subequations}
In the above results we have omitted some contributions of $T_{c11gg}$ and all of $T_{c11gk}$ and $T_{c11kk}$, because they are very lengthy. In fact, the expressions (\ref{c11gg}) are also extensive, however, it is worth considering them here, as we will see below. 

Finally, by taking into account the interchanges (\ref{Tc1234}), the first graph $T_{c1}$ takes the form:
\begin{eqnarray}\label{Tc1}
T_{c1}^{\mu_1\mu_2\mu_3\mu_4}(k_1,k_2,k_3,k_4) &=& T_{c1g\epsilon}^{\mu_1\mu_2\mu_3\mu_4}(k_1,k_2,k_3,k_4) + \sum_{i=1}^8T_{c1gg_i}^{\mu_1\mu_2\mu_3\mu_4}(k_1,k_2,k_3,k_4) \nonumber\\
&&+ \sum_{i=1}^{24}T_{c1gk_i}^{\mu_1\mu_2\mu_3\mu_4}(k_1,k_2,k_3,k_4) + \sum_{i=1}^{18}T_{c1kk_i}^{\mu_1\mu_2\mu_3\mu_4}(k_1,k_2,k_3,k_4),
\end{eqnarray}
where
\begin{eqnarray}
T_{c1g\epsilon}^{\mu_1\mu_2\mu_3\mu_4} &=& -\left[\frac{e^4}{6\pi^2\epsilon}-\frac{e^4}{12\pi^2}\ln\left(\frac{m^2}{\mu'^2}\right)\right]\nonumber\\
&&\times[2 g^{\mu_1\mu_4} c^{\mu_2\mu_3}+2 g^{\mu_1\mu_2} c^{\mu_3\mu_4}-4 g^{\mu_1\mu_3} c^{\mu_2\mu_4}\nonumber\\
&&+2  c^{\mu_1\mu_2} g^{\mu_3\mu_4}-4 g^{\mu_2\mu_4} c^{\mu_1\mu_3}+2g^{\mu_2\mu_3} c^{\mu_1\mu_4}\nonumber\\
&&-c_{\mu\nu}g^{\mu\nu}(g^{\mu_1\mu_4} g^{\mu_2\mu_3}-2 g^{\mu_1\mu_3} g^{\mu_2\mu_4}+g^{\mu_1\mu_2} g^{\mu_3\mu_4})]
\end{eqnarray}
and
\begin{equation}\label{Tc1gg}
\sum_{i=1}^4T_{c1gg_i}^{\mu_1\mu_2\mu_3\mu_4} = \sum_{i=1}^4T_{1gg_i}^{\mu_1\mu_2\mu_3\mu_4} \big |_{\{\kappa\to c_{\mu\nu}g^{\mu\nu},\kappa^2\to0,\tilde k_i\to k_i\}}.
\end{equation}
Thus, comparing the above perturbative expressions (\ref{Tc1gg}) with the nonperturbative ones (\ref{T1gg}) we observe that they are similar, being the results of Eq.~(\ref{Tc1}) an expansion up to first order in $c_{\mu\nu}$ of the equation (\ref{T1b}), with  $c_{00}=\kappa$ and $c_{0i}=c_{ij}=0$. Note that, as expected, the expression (\ref{Gcoll}) for the vacuum photon splitting is obtained from (\ref{Tc1gg}), when we consider the collinear limit.

Therefore, we can easily deduce the action coming from $T_{c1}$ (\ref{Tc1}) whose expression is in fact the perturbative Lorentz-violating Euler-Heisenberg action, previously obtained in~(\ref{SEH3}), nevertheless here, for a generic coefficient $c_{\mu\nu}$, given by
\begin{eqnarray}
S_{c\rm EH} &=& -\frac{\alpha^2}{90m^4} \int d^4x\, (k_F)_{\mu\nu\alpha\beta}(5F^{\alpha\beta}F^{\mu\nu}F_{\lambda\rho}F^{\lambda\rho} - 14 F^{\alpha\beta}F^{\nu\lambda}F_{\lambda\rho}F^{\rho\mu}) \nonumber\\
&&+\frac{\alpha^2}{180m^4} \int d^4x\, c_{\alpha\beta}g^{\alpha\beta}(5F_{\mu\nu}F^{\mu\nu}F_{\lambda\rho}F^{\lambda\rho} - 14F_{\mu\nu}F^{\nu\lambda}F_{\lambda\rho}F^{\rho\mu}),
\end{eqnarray}
where $(k_F)_{\mu\nu\alpha\beta}$ is defined in Eq.~(\ref{kF}). 

\section{Summary}\label{summary}

In this work, we have studied the radiative generation of the Lorentz-violating Euler-Heisenberg action, in the weak field approximation. Firstly, we have considered a nonperturbative calculation in the coefficient $c_{\mu\nu}$, however, by assuming rotational invariance, such that only $c_{00}\neq0$. The preliminary results are presented in Eq.~(\ref{T1b}), resulting then in the action (\ref{SEH2}). From these expressions, we are now able to calculate some scattering amplitudes relating to four-photon diagrams, in order to numerically estimate the coefficient for Lorentz violation. Within this approach, we have also recovered the results of the amplitude for the photon triple splitting (\ref{Gcoll}), previously obtained in the literature. Finally, we have taken into account the perturbative approach, where $c_{\mu\nu}$ is treated as a insertion in the propagator and a new vertex. The partial results are shown in Eq.~(\ref{Tc1}), which are in fact an expansion up to first order in $c_{\mu\nu}$ of Eq.~(\ref{T1b}), with $c_{00}=\kappa$ and $c_{0i}=c_{ij}=0$. This suggest that the complete results (\ref{T1b}) obtained in the nonperturbative approach can be used in both treatments. An extension of this work would be to consider higher-derivative terms, e.g., see the Lorentz-violating QED with operator of mass dimension six of Ref.~\cite{Rubtsov:2012kb}.

\vspace{.5cm}
{\bf Acknowledgements.} This work was supported by Conselho Nacional de Desenvolvimento Cient\'{\i}fico e Tecnol\'{o}gico (CNPq).

\end{document}